\documentstyle[preprint,aps]{revtex}

\newcommand{\bra}{\langle}
\newcommand{\ket}{\rangle}

\begin{document}
\draft
\title {State dependent effective interaction for the hyperspherical formalism}
\author{Nir Barnea$^{1,3}$, Winfried Leidemann$^{2,3}$,
and Giuseppina Orlandini$^{2,3}$}
\address{$^1$ECT$^*$, European Center for Theoretical Studies in Nuclear 
Physics and Related Areas, \\
Strada delle Tabarelle 286, 1-38050 Villazzano (Trento), Italy;\\
$^{2}$Dipartimento di Fisica, Universit\`a di Trento, I-38050 Povo, Italy;\\
$^{3}$Istituto Nazionale di Fisica Nucleare, Gruppo collegato di Trento.
}

\date{\today}
\maketitle

\begin{abstract}
The method of effective interaction, traditionally used in the framework
of an harmonic oscillator basis, is applied to the hyperspherical formalism 
of few--body nuclei ($A=3\div6$). The separation of the hyperradial part leads
to a state dependent effective potential. Undesirable features 
of the harmonic oscillator approach associated with the introduction of a 
spurious confining potential are avoided. It is shown that with the present 
method one obtains an enormous improvement of the convergence of the 
hyperspherical harmonics series in calculating ground state properties, 
excitation energies and transitions to continuum states. 
\end{abstract}

\section{Introduction}

Shell--model calculations in complex nuclei make
use of the single particle harmonic oscillator (HO) basis and 
are carried out in a truncated model space. In the last few years an 
impressive progress has been made in the application 
of these shell--model methods to the study of light nuclei 
~\cite{Zheng94,Zheng95,Navratil96a,Navratil98,Navratil99}. 
In the so called no--core shell--model calculations one keeps all the
nucleons active and, instead of the single particle HO basis, one introduces 
HO basis functions that depend on the Jacobi coordinates, thus removing
the spurious center of mass motion from the beginning 
~\cite{Navratil98,Navratil99}.

Since the HO series has a slow 
convergence rate one generally has to replace the bare nucleon--nucleon
interaction by an effective interaction tailored to the truncated model space.
Theoretically for a given model space one can find an effective interaction 
such that the spectrum of the effective A--body Hamiltonian coincides with a 
subset of the spectrum of the full--space bare Hamiltonian.
In practice, however, finding such an effective interaction is as difficult
as solving the full A--body problem. Therefore one resorts to an approximated
effective interaction, usually obtained from the solution of a 2--body 
Hamiltonian.  These 2--body effective interactions 
do no longer lead to the exact result in the truncated space,
but, if constructed properly, they retain two important properties:
(i) they converge to the bare Hamiltonian if the model space is enlarged
up to the full Hilbert space;
(ii) the energy levels of the effective Hamiltonian converge to the exact 
values faster than those of the bare Hamiltonian.

The HO basis functions resulting from a confining Hamiltonian do not posses
the correct asymptotic behavior of the nuclear A--body Hamiltonian. 
As a result the use of the HO basis may lead to a rather slow convergence 
of other observables besides the energy levels.
This limitation can be circumvented by using the hyperspherical harmonics (HH)
basis functions instead of the HO basis.
In the HH formalism, successfully applied to the nuclear few--body problem
~\cite{Fenin 72,Ballot80,Rosati92,BLO99},
the Jacobi coordinates are replaced by a single length
coordinate, the hyperradius, and a set of $3A-4$ hyperangles.
The HH are the A--body generalization of the 2--body spherical harmonics,
and likewise depend only on the hyperangular (angular) coordinates in the
hyperspherical (spherical) decomposition of the A--body (2--body) system.
In general, the wave function can be expanded in a series consisting of
products of HH basis functions and hyperradial basis functions. 
Very often the slow convergence rate of the HH basis is accelerated 
by using correlation functions ~\cite{Fenin 72,Rosati92,BLO99,Krivec90}. 
It is the aim of the present work to
investigate an alternative way of improving the convergence. To this end we 
reformulate the method of effective interaction for the HH expansion.

The HO and the HH expansion can become equivalent by making a particular 
choice for the hyperradial basis functions. Therefore  a trivial way to 
achieve a reformulation would be to make the HH expansion equivalent to 
the HO expansion. However, by doing so one would loose the extra 
flexibility the HH basis has in 
comparison to the HO basis and, moreover, impose an incorrect asymptotic 
behavior of the wave function. Therefore we renounce to such an
equivalent formulation. 

There is a further advantage of the HH basis, which is due
to the presence of the collective hyperradial coordinate. Eventually it will 
allow the introduction of a state dependent effective interaction similar to 
the recently formulated HO multi--valued effective interaction 
~\cite{Zheng95,Navratil96b}. 
At first sight, however, it appears that the hyperradius is leading to two 
problems. First, its collective feature seems to make it difficult to  single 
out a 2--body Hamiltonian in a natural way. Second, using 
a general form for the hyperradial basis functions one may find it difficult to
identify a model space in accordance with a 2--body effective interaction.
In this work we propose to solve these problems by defining a model space
that consists of a complete hyperradial set and the set of HH functions
with generalized angular momentum quantum number $K \leq K_{max}$.
The effective interaction is then deduced from an hyperangular Hamiltonian 
associated with the 2--body problem.

In Sec. \ref{sec:HH} we review the method of hyperspherical coordinates
and of the HH expansion. In Sec. \ref{sec:Veff} we derive the effective 
interaction for the HH expansion. Numerical results are then
given in Sec. \ref{sec:num} and conclusions are drawn is Sec. \ref{sec:con}.

\section{The hyperspherical Harmonic functions}
\label{sec:HH}

To introduce the hyperspherical coordinates we start from the 
center--of--mass coordinate 
$\roarrow{ R}=\frac{1}{A}\sum_{i=1}^A \roarrow{ r}_i$
and the normalized reversed order $N=A-1$ Jacobi coordinates
\begin{eqnarray} \label{Jacobi}
  \roarrow{\eta}_1 & = & \sqrt{\frac{A-1}{A}}\Big(\roarrow{ r}_1 
                - \frac{1}{A-1}(\roarrow{ r}_2 + \roarrow{ r}_3 + \cdots
                + \roarrow{ r}_{A} )\Big)  \nonumber \\
  \roarrow{\eta}_2 & = & \sqrt{\frac{A-2}{A-1}}\Big(\roarrow{ r}_{2} 
               - \frac{1}{A-2}(\roarrow{ r}_3 + \roarrow{ r}_4 + \cdots
               + \roarrow{ r}_{A} )\Big)  \nonumber \\
  &\ldots&  \nonumber \\ 
  \roarrow{\eta}_{N} & = & \sqrt{\frac{1}{2}}\Big( \roarrow{ r}_{A-1}
                        - \roarrow{ r}_{A} \Big) 
\end{eqnarray} 
where the $j$'th particle is specified relative to the center of mass of
particles $j+1$ to $A$.
The Jacobi coordinate $\roarrow{\eta}_j$ consists of a radial 
coordinate $\eta_j$ and a pair of angular coordinates 
$\hat{\eta}_j \equiv (\theta_j,\, \phi_j)$. 

These coordinates are then transformed into the hyperangular coordinates
$\alpha_2,\ldots,\alpha_{N}$ through the relation
\begin{equation} \label{hyper_n}
  \sin \alpha_n =  \eta_n / \rho_{n} \; ,
\end{equation}
where
\begin{equation} 
\label{rho_n}
\rho_{n}^2 = \rho_{n-1}^2 + \eta_n^2 = \sum_{j=1}^n \eta_j^2 \; .
\end{equation}
For $n=N$ we also find the relation
\begin{equation} \label{i2j}
  \rho^2 \equiv \rho_{N}^2 = 
  \frac{1}{A}\sum_{i<j}^A (\roarrow{r}_i-\roarrow{r}_{j})^2 \; .
\end{equation}
Therefore, the hyperradial coordinate $\rho$ is symmetric with respect to 
permutations of the underlying single particle coordinates.

The $3N=3(A-1)$ internal coordinates for the $A$--particle system consist of 
the hyperradial coordinate $\rho \equiv \rho_{N}$, and the $3N-1$ 
``hyperangular'' coordinates 
$\Omega_{N} \equiv \{\hat{\eta}_1,\, \hat{\eta}_2,\, \cdots,\, 
\hat{\eta}_{N}, \alpha_2,\, \alpha_3,\, \cdots,\, \alpha_{N}\} $.
These coordinates depend on the set of Jacobi 
coordinates specified in Eq. (\ref{Jacobi}).

Using the hyperspherical coordinates, the Laplace operator for
$n$ Jacobi coordinates, $n=1 \ldots N$,
can be written as a sum of two terms 
\begin{equation} \label{Laplacen}
  \Delta_{n} = \frac{1}{\rho_{n}^{3n-1}}\frac{\partial}{\partial \rho_{n}} 
  {\rho_{n}^{3n-1}} \frac{\partial}{\partial \rho_{n}}
   - \frac{1}{\rho_{n}^2} \hat{K}_n^2 \; .
\end{equation}
The  hyperspherical, or grand angular momentum 
operator $\hat{K}_n^2$ of the $n$ Jacobi coordinates can be expressed in 
terms of $\hat{K}_{n-1}^2$ and $\hat{\ell}_n^2$ as follows ~\cite{Efros}
\begin{equation} \label{Kn}
 \hat{K}_n^2 = - \frac{\partial ^2}{\partial \alpha_n^2} +
 \frac{3n-6 - (3n-2) \cos (2\alpha_n)} {\sin(2\alpha_n)}
 \frac{\partial}{\partial \alpha_n} 
+\frac{1}{\cos^2 \alpha_n}\hat{K}_{n-1}^2
+ \frac{1}{\sin^2 \alpha_n} \hat{\ell}_n^2 \; , 
\end{equation}
where we define $\hat {K}_1^2 \equiv \hat{\ell}_1^2$. 
The angular momentum operator associated with these $n$ coordinates
is $\roarrow{\hat{L}}_n = \roarrow{\hat{L}}_{n-1} +\roarrow{\hat{\ell}}_n$.
The operators $\hat{K}_{n}^2$, $\hat{\ell}_{n}^2$, $\hat{K}_{n-1}^2$,
$\hat{L}_n^2$ and $\hat{L}_{n_{\mbox{\scriptsize $z$}}}$ 
commute with each other. 

The hyperspherical harmonics functions ${\cal Y}_{[K_n]}$ are the 
eigenfunctions of this hyperangular
operator. The explicit expression for the HH functions of the first $n$ 
Jacobi coordinates is given by ~\cite{Fabre83}
\begin{eqnarray} \label{HH}
 {\cal Y}_{[K_n]} & = & \big[ \sum_{ m_1, \ldots ,m_n }  
 \bra \ell_1 m_1  \ell_2 m_2 | L_2 M_2 \ket 
 \bra L_{2} M_{2} \ell_3 m_3 | L_3 M_3 \ket \times \ldots \nonumber \\ 
 & & \; \; \; \times
 \bra L_{n-1} M_{n-1} \ell_n m_n | L_n M_n \ket 
 \prod_{j=1}^{n} Y_{\ell_j, \, m_j} (\hat{\eta}_j) \big] \times
 \nonumber \\ & & 
\big[ \prod_{j=2}^{n} \bbox {\cal N}_j( K_j ; \ell_j K_{j-1} ) 
(\sin \alpha_j )^{\ell_j} (\cos \alpha_j )^{K_{j-1}} 
P_{\mu_j}^{( \ell_j + \frac{1}{2},K_{j-1} + \frac{3j-5}{2}) }
(\cos (2\alpha_j) ) \big] \; ,
\end{eqnarray}
where $Y_{\ell, \, m}$ are the spherical harmonic functions, $P_{\mu}^{(a,b)}$
are the Jacobi polynomials and ${\cal N}_j( K_j ; \ell_j K_{j-1})$ are 
normalization constants given by ~\cite{Efros}:
\begin{equation} \label{norN}
  \bbox {\cal N}_j(K_j ; \ell_j K_{j-1}) = \left[ \frac
  {(2K_j+3j-2)\mu_j!\Gamma(\mu_j+K_{j-1}+\ell_j+\frac{3j-2}{2})}
  {\Gamma(\mu_j+\ell_j+\frac{3}{2})
  \Gamma(\mu_j+K_{j-1}+\frac{3j-3}{2})} 
  \right]^{\frac{1}{2} } \; .
\end{equation}
The symbol $[K_n]$ stands for the set of quantum numbers $\ell_1,...,\ell_n$,
$L_2,...,L_n$, $\mu_2,...,\mu_n$ and $M_n$.
The quantum numbers $K_j$ are given by
\begin{equation} \label{K_N}
K_j = 2\mu_j + K_{j-1}+ \ell_j \,\,\,\, ; \mu_1 \equiv 0 \; ,
\end{equation}
and the $\mu_j$ are non--negative integers.
By construction, $\rho_n^{K_n} {\cal Y}_{[K_n]}$ is an harmonic polynomial
of degree $K_n$. The HH function ${\cal Y}_{[K_n]}$ is an eigenfunction of
$\hat{K}_n^2$ with eigenvalues
\begin{equation} \label{Nvalue}
  K_n(K_n + 3n - 2) \;.
\end{equation}

It is evident that the HH functions (\ref{HH}) do not posses any special
properties under particle permutation. Therefore the first step in 
applying the HH expansion to the A--body problem is the symmetrization
of the HH basis. In the current work we  employ two powerful algorithms
~\cite{Nir9798,Akiva94} recently developed for the construction 
of an HH basis with
well defined permutational symmetry. This enabled us to check our results,
since they could be obtained in two independent ways.

In view of Eqs. (\ref{rho_n}) and (\ref{i2j}) it is evident that the HO 
Hamiltonian,
written in the form
\begin{equation}\label{HHO}
\sum_{j=1}^N\left(-{\Delta_{j}\over 2}+\omega^2 \eta_j^2\right)={1\over 2}\left
({\partial^2\over\partial\rho^2}+{3N+4\over \rho}{\partial\over\partial \rho}
-{\hat K^2\over \rho^2}+\omega^2\rho\right) \; ,
\end{equation}
has eigenvectors of the following form
\begin{equation}\label{psiHO}
\Psi_{HO}=R_{n_\rho}(\rho)\,{\cal Y}_{[K]}
\end{equation}
with eigenvalues
\begin{equation}\label{eigenHO}
E_n=\hbar\omega\left({3(A-1)\over2}+n\right)=\hbar\omega\left(
{3(A-1)\over2}+2n_\rho+K\right)\,.
\end{equation}
Therefore the HH $K$--quantum number can be associated with the $n=K$
excitations of the HO wave function.

\section{The effective interaction}
\label{sec:Veff}

In general we would like to use the HH basis functions to solve the
A--body Hamiltonian
\begin{equation}\label{H_0}
  H = \sum_{i=1}^{A} \frac{\vec{p}_i^{\,2}}{2 m} + \sum_{i<j}^{A} V_{ij} \;,
\end{equation}
where $m$ is the nucleon mass and $V_{ij}$ is the nucleon--nucleon (NN)
interaction. In practice, looking for the eigenvectors of $H$ in terms of
the HH expansion turns out to be a notoriously difficult task. Therefore,
one usually has to introduce correlation functions in order to accelerate
the convergence of the calculation ~\cite{Fenin 72,Rosati92,BLO99,Krivec90}. 
In this work, however, we shall explore
another possibility and instead of using correlation functions we shall use 
the method of effective interaction ~\cite{Suzuki80}. This approach
is largely used in shell--model calculations (see e.g. Ref.~\cite{Morten}), 
where the 
harmonic oscillator basis is used in a truncated model space. Instead of the
bare nucleon--nucleon interaction one uses effective interactions inside the
model space. Defining $P$ as the projection operator into the model space and 
$Q=1-P$ as the projection into the complementary space, the model space
Hamiltonian can be written as
\begin{equation}\label{H_P}
  H_P = P\left[ \sum_{i=1}^{A} \frac{\vec{p}_i^{\,2}}{2 m} \right] P + 
        P \left[ \sum_{i<j}^{A} V_{ij} \right]_{eff} P \;.
\end{equation}
In general the effective interaction appearing in Eq. (\ref{H_P}) is an A--body
interaction. If it is determined without any approximation, the model--space 
Hamiltonian provides a set of eigenvalues which coincides with a subset of the 
eigenvalues of the original full--space Hamiltonian, Eq. (\ref{H_0}).
However, calculation of the exact A--body effective interaction is as 
difficult as finding the full--space solution. 

In the HH formalism the model
space can be defined as a product of the hyperradial subspace and the complete
set of HH basis functions with generalized angular momentum quantum number 
$K \leq K_{max}$. Instead of 
calculating the exact effective interaction we shall look for an approximate
effective interaction with the following properties:
\begin{itemize}
\item $V_{eff} \longrightarrow V$ as $K_{max} \longrightarrow \infty$;
\item the eigenvalues, $E_i(K_{max})$, and eigenvectors of the effective 
A--body Hamiltonian converge to their limiting values faster than the 
eigenvalues and eigenvectors of the bare Hamiltonian.
\end{itemize}
Let us now turn to the problem of constructing the effective interaction.
It is customary to approximate $V_{eff}$ by a sum of two--body effective 
interactions determined from a 2--body problem. 
As the nuclear 2--body system contains only one bound state one is forced
to introduce a confining potential into the 2--body problem
in order to ensure large overlaps between the model space states and the 
eigenvectors of the 2--body problem. It will be shown that in the present 
approach one does not need such an additional confining potential.
 
Using the symmetrized HH basis the matrix elements of the effective 
interaction can be deduced from the matrix elements of the ``last'' pair,
\begin{equation}
  \bra \sum_{i<j}^A V_{2\;eff}(\vec{r}_{ij}) \ket = 
  \frac{A(A-1)}{2}\bra V_{2\;eff}(\vec{r}_{A,A-1}) \ket \;.
\end{equation} 
The relevant hyperspherical degrees of freedom 
associated with $V_{2\;eff}(\vec{r}_{A,A-1})$ are $\hat{\eta}_N$
and the hyperangle,
\begin{equation}
	\sin \alpha_N = \frac{r_{A,A-1}}{\sqrt{2}\rho} \;.
\end{equation}
A natural choice for the corresponding hyperspherical ``2--body'' 
Hamiltonian is
\begin{equation}\label{H2}
  H_2(\rho) = \frac{1}{2 m} \frac{\hat{K}_N^2}{\rho^2} 
            + V(\sqrt{2}\rho \sin \alpha_N \cdot \hat {\eta}_N) \;,
\end{equation}
since $\hat{K}_N^2$ contains the canonical kinetic energy associated
with the two--body variables $\alpha_N$ and $\hat{\eta}_N$ (see Eq. (\ref{Kn})).
Such an $H_2$ is in fact an A--body effective interaction
as it contains the hyperspherical part of the A--body kinetic energy
operator and it is a function of the collective coordinate $\rho$.
The hyperradial kinetic energy operator has not been included in $H_2$. 
The reason is that we can use a complete basis set for the $\rho$--space
and therefore we do not need  to
define an effective interaction for the hyperradial part. 

Due to the collective coordinate, $\rho$, in $H_2$ one has automatically  a 
confinement of the 2--body--system: for moderate values of $\rho$
the relation $0\leq r_{A,A-1}\leq\sqrt{2}\rho$ ensures localization of the
2--body wave function and for large values of $\rho$ the effective Hamiltonian
coincides with the bare one, since the NN interaction vanishes. Therefore 
large overlaps between the model space 
states and the eigenvectors of the 2--body problem are ensured. 

Different from the HO approach 
we do not calculate an effective interaction for a "free" 2--body--system 
but for a "bound" one. Therefore we can avoid to 
introduce in $H_2$ the additional confining potential which is necessary 
for the HO effective interaction and which leads to undesirable features.

The matrix elements of $H_2$ between the A--body HH functions, Eq. (\ref{HH}),
are given by
\begin{eqnarray}
  \bra [K_N] | H_2(\rho) | [K'_N] \ket & = &
   \delta_{[K_N][K'_N]}\frac{1}{2 m} \frac{K_N(K_N+3N-2)}{\rho^2} \cr
  & + &\delta_{[K_{N-1}][K'_{N-1}]}
   V^{K_{N-1}L_{N-1}}_{K_N L_N \ell_N, K'_N L'_N \ell'_N}(\rho) \;,
\end{eqnarray}
where 
\begin{equation}
V^{K_{N-1}L_{N-1}}_{K_N L_N \ell_N, K'_N L'_N \ell'_N}(\rho)=
   \int d \Omega_N {\cal Y}_{[K_{N}]}^* 
                   V(\sqrt{2} \rho \sin \alpha_{N}\cdot \hat{\eta}_N)
                   {\cal Y}_{[K_{N}']} \; .	
\end{equation}
We see that $H_2$ is diagonal in the quantum numbers $[K_{N-1}]$ and for 
central potentials also in $L_N, \ell_N$. Due to the hyperangular integration 
$H_2$ explicitly depends on quantum numbers of the residual system, i.e.
$K_{N-1}$ and for non-central forces also $L_{N-1}$. The effective 2--body
Hamiltonian is independent of the other quantum numbers in $[K_{N-1}]$
(see Eq. \ref{HH}). As a result the HH 
effective interaction depends on the state of the residual $A-2$ particle
subsystem. Such a "medium correction" of the 2--body force is of course a great 
advantage of our approach and is similar to the HO multi--valued effective 
interaction ~\cite{Zheng95,Navratil96a}.
On the other hand one has to pay for it with a greater numerical effort, since
the effective interaction has to be calculated for all the various states
and does also depend on the specific $A$--body system.

Solving the hyperradial equation on a 
grid, we have to diagonalize $H_2$ for each grid point $\rho_i$ and for all
the possible values of $K_{N-1}$ in our model space. In general one should 
reach a $K_{\rm MAX}$ value for the Q-space around 60 for the ground state of 
the s-shell nuclei and $K_{\rm MAX}\sim 200$ for p-shell nuclei and excited states. 
>From this point 
we can follow the same procedure as Barrett and Navratil
~\cite{Navratil98,Navratil99}.
Employing the Lee--Suzuki ~\cite{Suzuki80} similarity transformation method, 
we can use the
eigenvectors, $\{|i\ket\}$, and eigenvalues, $\{\epsilon_i\}$,
of $H_2^{K_{N-1},L_{N-1}}(\rho)$ to construct the effective interaction.
Let us denote by $|\alpha\ket$ the HH functions that belong to our 
model space,
i.e. the HH function $|[K_N]\ket$ such that $K_N \leq K_{max}$, and by 
$|\beta\ket$ the states that belong to the $Q$ space, 
$Q=\{ |[K_N]\ket\;;\; K_N > K_{max} \}$.
The Lee--Suzuki effective interaction then takes the form,
\begin{equation}
  P \tilde{H}_{2} P = P H_2 P + P H_2 Q \omega P \; , 
\end{equation}
where the transformation operator $\omega=Q \omega P$ is given by the
equation
\begin{equation}\label{omega}
  \bra \beta | i \ket = \sum_{\alpha} \bra \beta | \omega | 
\alpha \ket \bra \alpha | i\ket \;.
\end{equation} 
If $n_P$ is the number of model--space HH basis functions that belong to the 
subspace $K_{N-1}$, we may solve Eq. (\ref{omega}) for $\omega$ by choosing 
a set, $\cal{A}$, of $n_P$ eigenvectors with the lowest eigenvalues $|i\ket$ 
and inverting the matrix $\bra \alpha | i\ket$.
The resulting effective 2--body Hamiltonian 
\begin{equation}\label{H2tilde}
 \bra \alpha | \tilde{H_2}(K_{N-1},\rho) | \alpha'\ket = 
  \sum_i^{n_P} \left[ 
       \bra \alpha | i \ket \epsilon_i \bra i | \alpha' \ket +
       \sum_{\beta} \bra\alpha | i\ket \epsilon_i \bra i|\beta \ket
       \bra \beta | \omega | \alpha \ket \right]
   \;,
\end{equation} 
will have the property that $P|i\ket$, $|i\ket\in\cal{A}$, is a right 
eigenvector of 
$\tilde{H_2}$ with eigenvalue $\epsilon_i$.
The effective interaction is in general a non--hermitian operator, however
it can be hermitized, using the transformation ~\cite{Suzuki82}
\begin{equation}\label{H2eff}
  H_{2\;eff} = [P(1+\omega^{\dagger}\omega)P]^{1/2}\tilde{H_2}
               [P(1+\omega^{\dagger}\omega)P]^{-1/2} \;.
\end{equation}
The effective interaction can be now deduced from $H_{2\;eff}$, by 
subtracting the kinetic energy term,
\begin{equation}\label{Veff}
  V_{eff} = H_{2\;eff} - \frac{1}{2 m} \frac{\hat{K}_N^2}{\rho^2} \;.
\end{equation}
It can be seen that as $K_{max}\longrightarrow \infty$ $V_{2eff}$ indeed 
reproduces the bare NN interaction in contrast to the HO effective interaction,
where only the total Hamiltonian converges to the correct result. Another 
interesting feature of the current formulation is that, in contrast to HO 
effective interaction, the HH effective interaction vanishes at large distances 
as it should be for a system of non--interacting particles. In addition we 
would like to mention that similarly to the HO approach \cite{Navratil98,NB99} 
the present formalism can be extended to incorporate, beyond two--body, 
also three-- and more--body effective interactions.
\section{Numerical results}
\label{sec:num}

In order to check the proposed formulation of the hyperspherical
effective interaction, we apply the described formalism to few--body nuclei
in the mass range $A = 3\div6$. The following simple nucleon--nucleon
interactions are used: Malfliet--Tjon potentials MTV ~\cite{MTV} and MT-I/III 
~\cite{MTI-III} as well as the Minnesota potential MN ~\cite{MN} (see Table 1).
Of course we are aware that realistic potential models have already been used
even for nuclei with $A>4$ \cite{argonne}, but the principal aim of the present work 
is the introduction of the HH effective interaction.
To this end we investigate the rate of convergence of the HH series with the 
effective interaction for ground state energies, radii, and excitation 
energies. In addition we study transitions to continuum states via the method 
of Lorentz integral transform ~\cite{ELO94}. 

Expanding the effective wave functions into 
HH basis functions the effective Hamiltonian (Eq. (\ref{H_P})) is transformed 
into a set of coupled differential equations in the hyperradius $\rho$. 
These equations are then solved expanding the solution in terms of generalized 
Laguerre polynomials.

Our results for ground state energies and radii as well as first excitation
energies are summarized in Table II. In case that results from other authors 
were available they are also given in the table. In general one observes 
a very good agreement comparing the results of the various methods. In the 
following the quality of the convergence for the calculated
observables is discussed in detail.

In Fig. 1 we illustrate the convergence patterns with bare and effective
interactions for binding energy and radius of the $A=3$ system with the MN 
potential. It is readily seen that the effective interaction improves the
convergence drastically. Already with $K=2$ one finds for the energy (radius) 
a deviation of only 0.6 \% (1.3 \%) from the converged value, while with the 
bare interaction one 
needs $K=10 (8)$ for a similarly good result. For $K=10$ one obtains  
sufficiently converged results with the effective interaction, whereas for the
bare interaction one has to go up to $K=18$ or higher to reach a similar 
precision.

In Fig. 2 we show the corresponding results for $^4$He with the MTV potential.
In this case the difference between the convergence of effective and 
bare interactions is even stronger. 
For the effective interaction one obtains almost the correct values for
energy and radius with a rather low $K$ of 4. The convergence with the bare
interaction is considerably worse, even with $K=20$ one does not have 
completely converged results. In Fig. 3 we compare our results to those
of Navratil and Barrett with the HO effective interaction ~\cite{NB99}. 
One obtains also for the HO case a very nice convergence, but it depends
on the chosen harmonic oscillator frequency $\Omega$. On the other
hand it is evident that the parameter free HH effective interaction leads 
to a considerable improvement.

In Figs. 4-6 we give an overview of our results for the bound state
properties with the various potential models. For $A=4$ one finds a very
good convergence both for binding energy and radius. For the systems with
$A>4$, which approximately can be considered as an $\alpha$--core plus
remaining nucleons at larger distances, there is generally a good  
convergence, but for the case of the radii with the MTV potential. 
However, we do not reach the same extremely good precision as 
for the cases with $A \le 4$, since we have to restrict our calculation to 
a somewhat lower $K$. This is due to the fact that an increasing number of 
nucleons leads to a much higher number of HH functions for the same $K$.
In general our calculations are limited to about 400 HH basis functions.

The first excitation energies of the $A=6$ systems 
are illustrated in Fig. 7. One sees that the convergence patterns are
quite similar. It shows that also non ground state observables can be
calculated with sufficient precision with the HH effective interaction.

One may ask what happens at even higher energies, e.g. in reactions where
states in the continuum are involved. In order to address this
question we consider as next point
transitions from the ground state to continuum states due
to an external probe. Such response functions can be calculated with the
method of the Lorentz integral transform ~\cite{ELO94}. In this formalism,
besides the ground state, one needs an additional ``Lorentz--state''
$\tilde\Psi$, which is localized and 
which carries all the information about the excitation of the system and 
of the final state interactions. The convergence of the Lorentz integral 
transform depends on the convergence on the hyperangular quantum number
of the ground state ($K_0$) and that of
$\tilde\Psi$ ($K_T$). Both effects are illustrated in Fig. 8, where we show the 
Lorentz integral transform of the $^4$He total photoabsorption cross section 
in the electric dipole approximation with the MT-I/III potential. One sees
that for a convergent result of the transform a rather low $K_T$ for
the HH expansion of $\tilde\Psi$ is sufficient, while one has to include 
somewhat higher $K_0$'s in the HH expansion of the ground state. 
Our result for the Lorentz transform is a bit lower than that of 
Ref. ~\cite{ELO97} since there, different from the Lorentz state, the ground
state was calculated without correlations and  was not
fully convergent, but the conclusions of Ref. ~\cite{ELO97} remain unchanged.

\section{conclusion}
\label{sec:con}

In this work we have introduced an hyperspherical effective interaction. 
To this end we have defined a model space consisting 
of a complete hyperradial basis and a set of HH functions with generalized 
angular momentum quantum number $K \le K_{max}$. The effective interaction
has been derived from an hyperangular Hamiltonian connected with the two--body 
problem. The Hamiltonian also includes the hyperangular kinetic energy, which
is proportional to 1/$\rho^2$, where $\rho$ is the collective hyperradial
coordinate. Because of this additional $A$--body piece the $A-2$ residual 
system cannot be considered as a pure spectator. It leads to an 
effective interaction depending explicitly on the state of the residual system,
similar to the HO multi--valued effective interaction.

We should mention that the present approach can be extended in a 
straightforward way to derive an HH 3-- or more--body effective interaction.

We have applied the developed formalism to few--body systems in the mass range 
$A=3\div6$ calculating binding/excitation energies, radii and, via the method 
of the Lorentz integral transform, also reactions at energies far in the 
continuum. For the larger $A$'s these calculations have become feasible due to 
a powerful algorithm to construct symmetrized basis states. In general we 
obtain nicely converging results showing that the HH 
effective interaction is a very powerful tool. Particularly interesting 
is the fact that the rate of convergence is always very good and does not 
depend much on the considered observable. We believe that the inherent 
confinement of the effective ''two--body`` Hamiltonian is largely responsible
for this fact.

Where available we have compared our results to other calculations in the 
literature. In general a very good agreement is obtained. 

Eventually we would also like to point out that the present approach has 
various advantages compared to the harmonic oscillator formalism:
(i) one does not need to introduce an additional confining potential and 
contrary to the HO approach one obtains a parameter free effective interaction. 
Therefore one does not have a problem that is typical for the HO formalism, 
namely that the convergence of different observables, e.g. binding energies and 
radii, lead to rather different best choices for such a parameter;
(ii) the HH effective interaction is automatically state dependent; 
(iii) the HH basis functions are 
much better suited than the HO basis functions to describe the asymptotic part 
of the wave function. On the other hand it will require additional effort to
elevate the present approach to the same level of sophistication as the HO
formalism regarding the use of modern realistic interactions and number of
particles.

\acknowledgments
The authors would like to thank Petr Navratil for triggering this work 
by presenting his results at an ECT$^*$ workshop, for stimulating discussions and for providing us with the $^4$He results for the MTV potential. We
would also like to acknowledge Bruce Barrett,
and Kalman Varga for useful discussions.   


\begin{table} 
\caption{
\label{tb:pot_par}
List of the parameters of the N-N potential used in this work. The potential 
is written as a sum of a few terms; each is expressed as 
$V_i f(\mu_i,r)(W_i+B_i P_\sigma-H_iP_\tau+M_iP_r)$, 
where $P_\sigma\,,P_\tau\,, P_r$ are spin--, isospin--
and space--exchange operators. $f(\mu_i,r)=\exp{(-\mu r^2)}$ for Gauss--type
potential, $f(\mu_i,r)=\exp{(-\mu r)}/r$ Yukawa--type. The potential strengths
$V_i$ are in [MeV]. The range $\mu$ is in [fm$^{-2}$] for Gauss--type or 
[fm$^{-1}$] for Yukawa type.}

\begin{center}
\begin{tabular}
{lccccccccc} 
    Potential &  Type & i &  $V_i$  & $\mu_i$ & $W_i$ & $M_i$ & $B_i$ & $H_i$
\\ \hline
MTV ~\cite{MTV}& Yukawa& 1 & 1458.05 &  3.11   &  1.0  &  0.0  &  0.0  & 0.0 
\\
              &       & 2 &- 578.09 &  1.55   &  1.0  &  0.0  &  0.0  & 0.0
\\ \hline
MTI-III ~\cite{MTI-III}& Yukawa& 1 &  1458.27&  3.11   &  0.5  &  0.5  &  0.0  
& 0.0
\\
              &       & 2 & -578.178&  1.555  & 0.5&  0.5  &  0.0495  &0.0495
\\ \hline 
 MN ~\cite{MN}& Gauss & 1 &  200.0  &  1.487  &  0.5  &  0.5  &  0.0  & 0.0 
\\
              &       & 2 & -178.0  &  0.639  &  0.25 &  0.25 &  0.25 & 0.25
\\
              &       & 3 &  -91.85 &  0.465  &  0.25 &  0.25 & -0.25 &-0.25
\\ \hline

\end{tabular}
\end{center}
\end{table}

\begin{table} 
\caption{\label{tb:asy} 
         Comparison of binding energies ($E_B$) in [MeV] and root mean square 
radii ($<r^2>^{1/2}$) in [fm] obtained with the present method of effective 
interaction in the HH formalism (EIHH) with results of other methods. 
In the case of EIHH the calculations with MTI--III and MN 
potentials include the Coulomb interaction. For EIHH
the number in parenthesis indicates the variance with respect to the result
obtained with $K=K_{max}-1$. The quality of the convergence can be inferred 
from Figs. 1-6.}

\begin{center}
\begin{tabular}
{lccccccc} 
                  & &MN& &MTI--III& &MTV\\ \hline 
Nucleus&Method [Ref.]&$E$&$<r^2>^{1/2}$&$E$&$<r^2>^{1/2}$&$E$&$<r^2>^{1/2}$
\\ \hline 
\\
$^3$H          & EIHH    & -8.3856(5) & 1.7036(1) & -8.718(9) & 1.7064(2) & 
-8.244(8)  & 1.6798(3)
\\
  &EIHO ~\cite{Navrprivcom}    &   --    &    --     &    --     &   --      &
-8.235(5)    &  --
\\
  &SVM ~\cite{Varga95}    & -8.380     & 1.698     &    --     &   --      &
-8.2527    & 1.682
\\
  &Faddeev ~\cite{Friar81}& --         &   --      &  -8.54    &   --      & 
-8.25273   &  --
\\
& ATMS ~\cite{Akaishi86}  & --         &   --      &    --     &   --      & -8.26(1)   & 1.682 
\\
 & CHH1 ~\cite{Kievsky}   & --        &   --      &    --     &   --      & 
-8.240     &  --
\\ 
  & CHH2  ~\cite{ELO97b}   & --        &   --      &  -8.74    &   --      & 
--     &  --
\\ 
 & GFMC ~\cite{Zabol}     & --         &   --      &    --     &   --      & 
-8.26(1)   & 1.682
\\
   & VMC ~\cite{Varga95}  & --         &   --      &    --     &   --      & 
-8.27(3)   & 1.68
\\
\\
$^3$H$^\star$  & EIHH    & -0.421(9)  & 4.757(5)  & -1.01(1) & 7.2(1)    & 
-0.073(6) & 6.973(8)
\\
\\ \hline

$^4$He         & EIHH    & -29.96(1)  & 1.4106(1) & -30.71(2) & 1.4222(2) & -31.358(9) & 1.40851(3)
\\
  & SVM ~\cite{Varga95}   & -29.937    & 1.41      &    --     &   --      & 
-31.360    & 1.4087
\\
   & FY ~\cite{gloeckle}    & --         &   --      &  -30.29   &   --      &
-31.36     &  --
\\
   & ATMS ~\cite{Akaishi86}  & --         &   --      &    --     &   --      & 
-31.36     & 1.40 
\\
   & CHH2  & --         &   --      &  -30.69   &   1.421     & 
--     &  --
\\
 & CRCG ~\cite{Kamimura90}  & --         &   --      &    --     &   --      & 
-31.357     &  --
\\ 
  & GFMC ~\cite{Zabol}  & --         &   --      &    --     &   --      & 
-31.3(2)   & 1.36
\\
  & VMC ~\cite{Varga95}   & --         &   --      &  --       &   --      &
-31.30(5)  & 1.39
\\
\\
$^4$He$^\star$ & EIHH    & -7.848(4)  & 3.405(7)  & -8.48(1)  & 3.59(2)   &
    --     &  -- 
\\
\\ \hline

$^5$He         & EIHH    & -28.1(2)   & 2.17(6)   & -29.4(1)  & 2.13(5)   & 
-43.7(3)  & 1.499(5) 
\\
               & SVM ~\cite{Varga95}     & --         & --        &    --     &    --     & 
-43.48     & 1.51
\\ 
               & VMC ~\cite{Varga95}   & --         &   --      &    --     &   --      &
-43.0(2)   & 1.51
\\
\\
$^5$He$^\star$ & EIHH    & -20.8(9)   & 3.33(1)   & -22.(1)   & 3.312(2)  &
   --      &  -- 
\\
\\ \hline

$^6$He         & EIHH    & -30.6(5)   & 2.323(2)  & -32.5(3)  & 2.263(9)  & 
-68.5(2)   & 1.512(4)
\\
               & SVM ~\cite{Varga95}     & -30.07     & 2.44      &    --     &   --      & -66.30     & 1.52
\\
               & VMC ~\cite{Varga95}     & --         & --        &    --     &   --      & -66.3(3)   & 1.50
\\
\\
$^6$He$^\star$ & EIHH    & -22.5(3)   & 3.55(9)   & -23.5(2)  & 3.54(3)   &
     --    &  --
\\ 
\\ \hline
 
$^6$Li         & EIHH    & -35.2(4)   & 2.16(2)   & -36.6(3)  & 2.15(2)     & 
-68.5(2)   & 1.512(4)
\\
& SVM ~\cite{Varga95}     & -34.59     &    2.22   &    --     &    --     &
     -66.30 &  1.52
\\ 
\\
$^6$Li$^\star$ & EIHH & -25.7(4)      & --        & -26.6(5)  & 3.35(3)   & --     & --
\\
\\ \hline
    
\end{tabular}
\end{center}
\end{table} 
\newpage

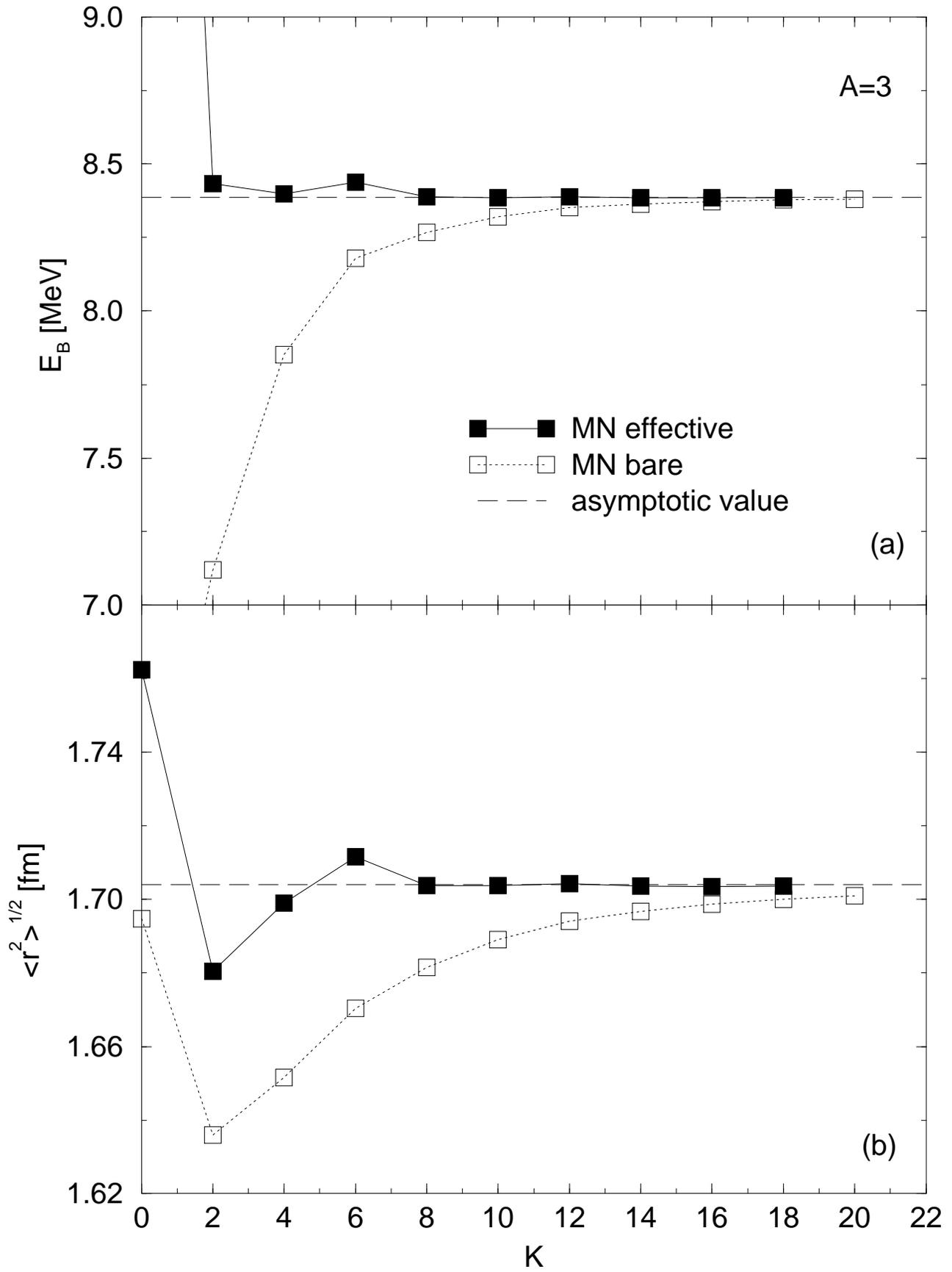
\begin{figure}
\caption
{Binding energy (a) and root mean square radius (b) of the A=3
system for Minnesota potential [22]
as a function of the hyperangular
quantum number $K$. The asymptotic value has  been indicated by a dashed 
line.}
\end{figure}

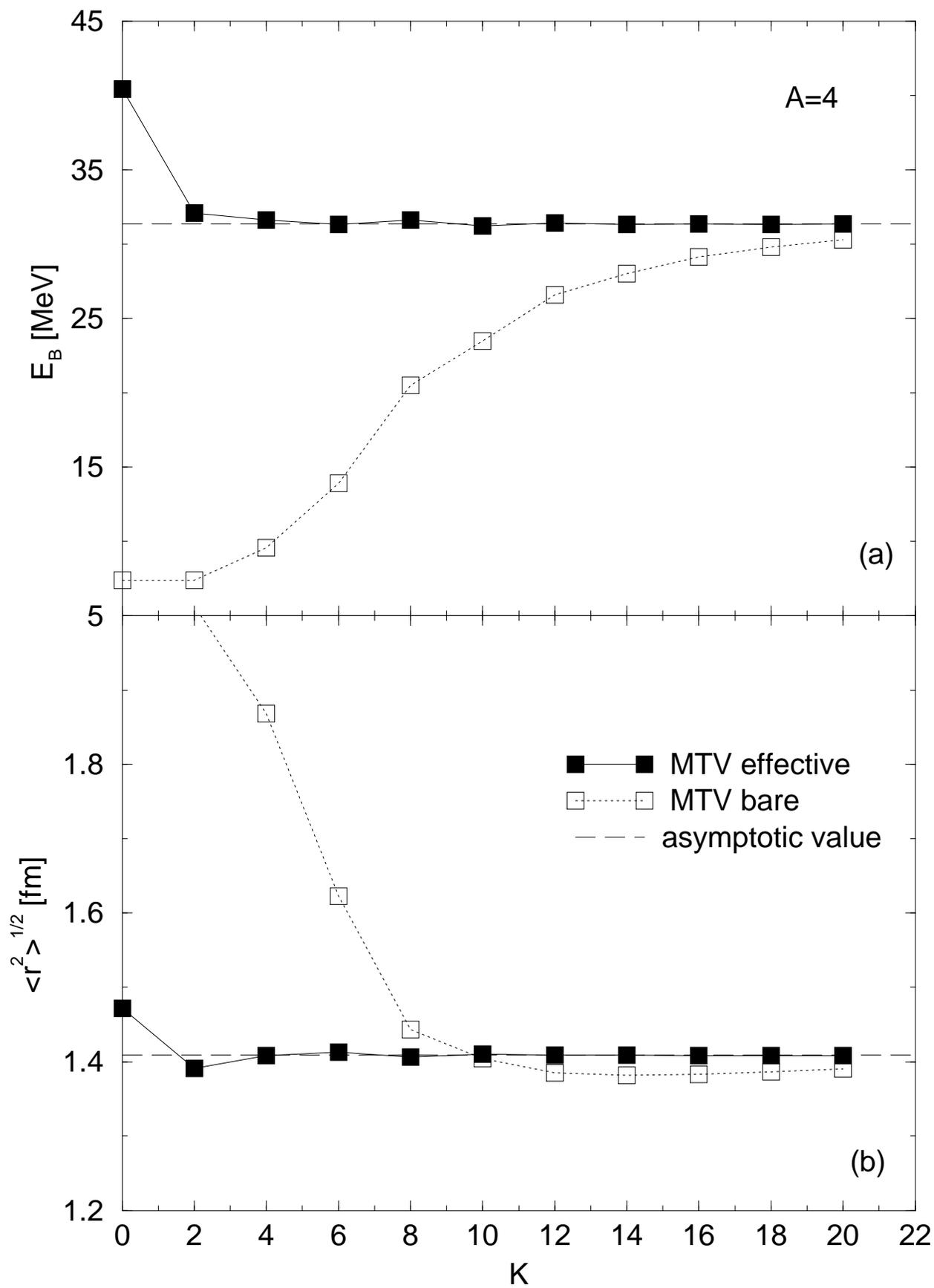
\begin{figure}
\caption{Same as Fig.1 for A=4 and the MTV potential [20]
}
\end{figure}

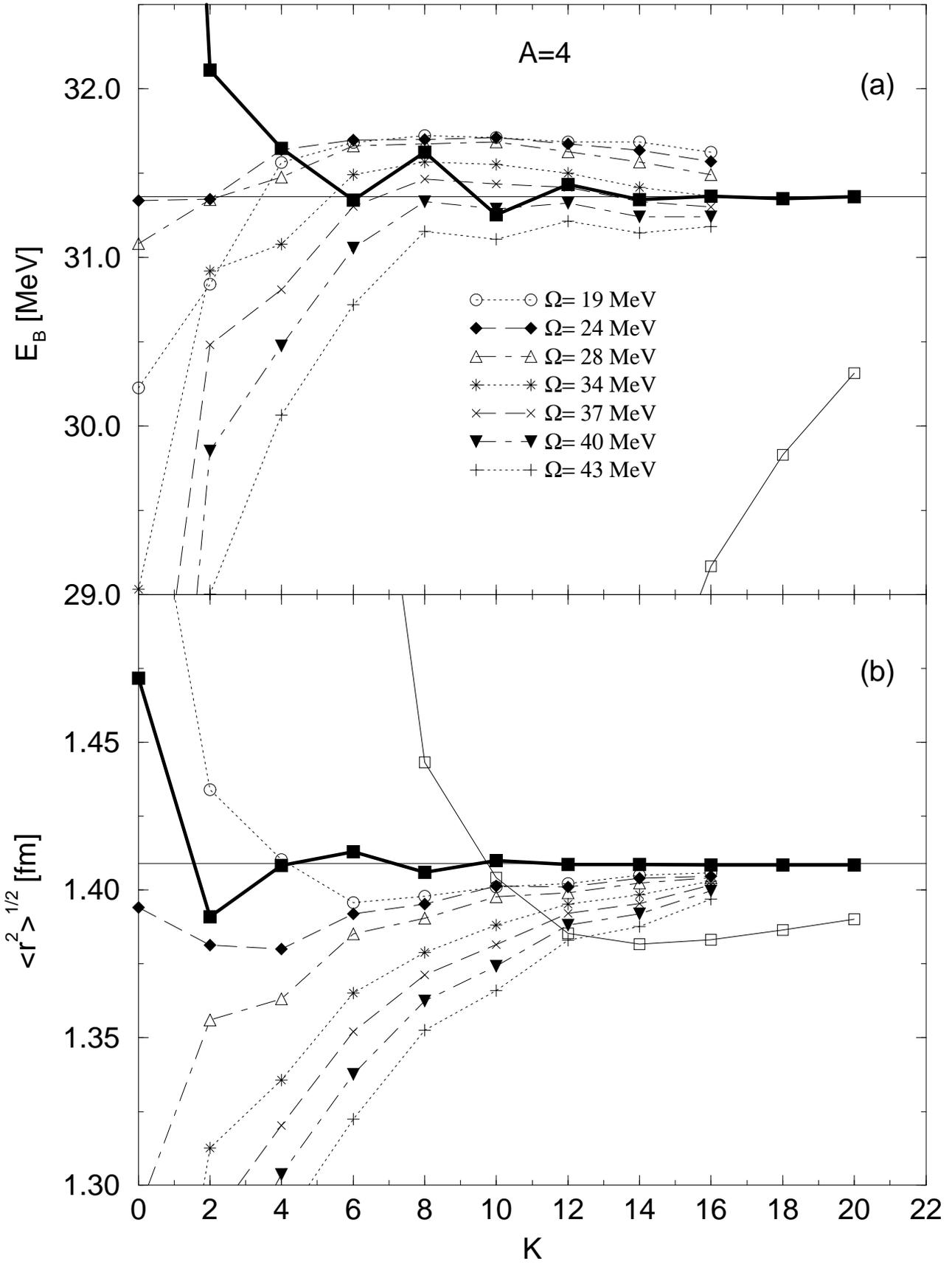
\begin{figure}
\caption{Comparison between results of the
present method (full squares: effective interaction, open squares: bare interaction) and that of Refs. [5,25] 
obtained with different values of the HO parameter $\Omega (\hbar =1)$. 
Binding energy (a) and root mean square radius (b) 
of the A=4 system for the MTV potential [20]
as a function of the hyperangular
quantum number $K$ or of the HO quantum number $N$.}
\end{figure}

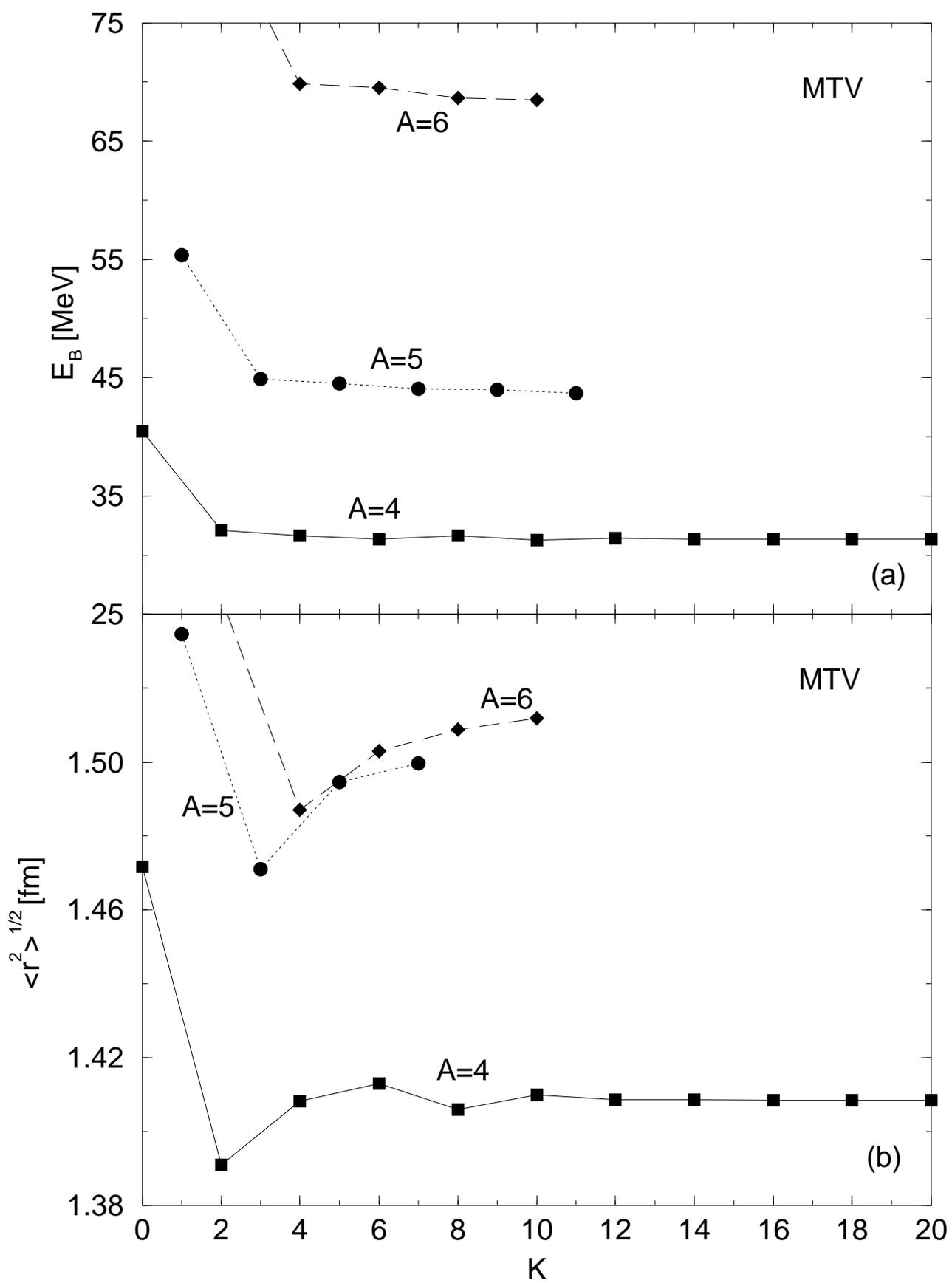
\begin{figure}
\caption{Binding energies (a) and root mean square radii (b) of different 
A-body systems for MTV potential [20] 
as a function of the hyperangular quantum number $K$.}
\end{figure}

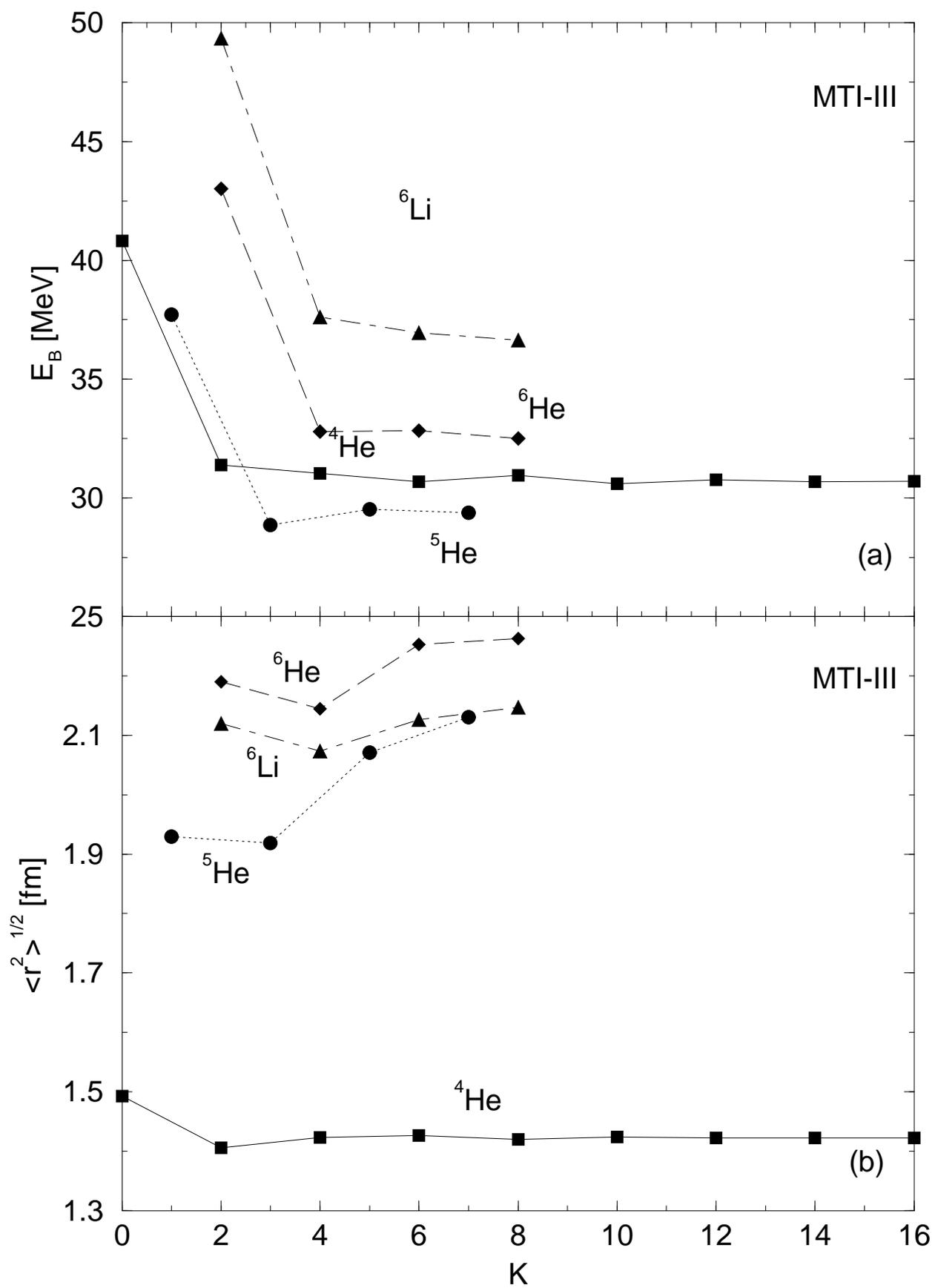
\begin{figure}
\caption{Same as Fig. 3 for MTI-III potential [21]
}
\end{figure}

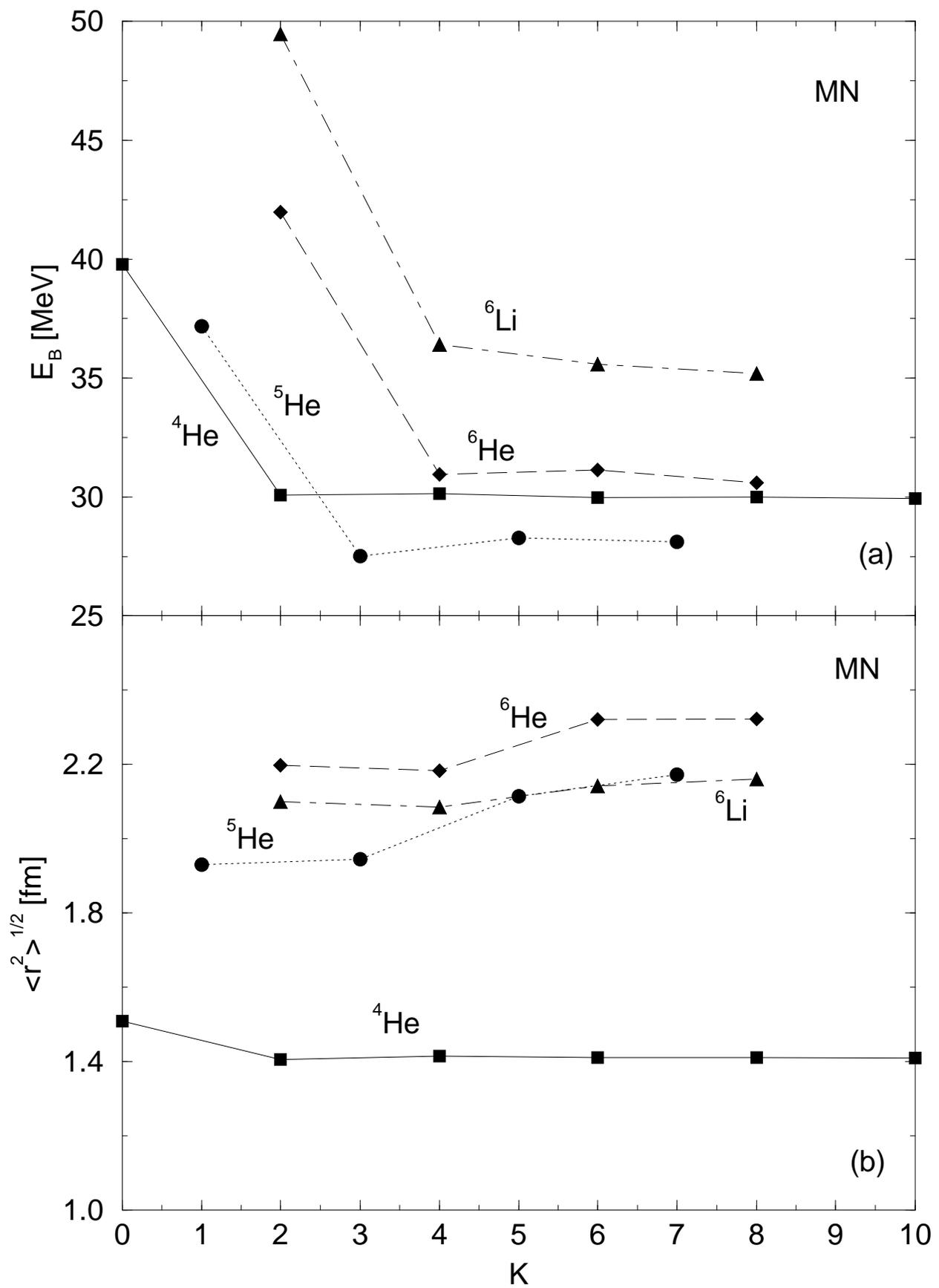
\begin{figure}
\caption{Same as Fig. 3 for MN potential [22]
}
\end{figure}

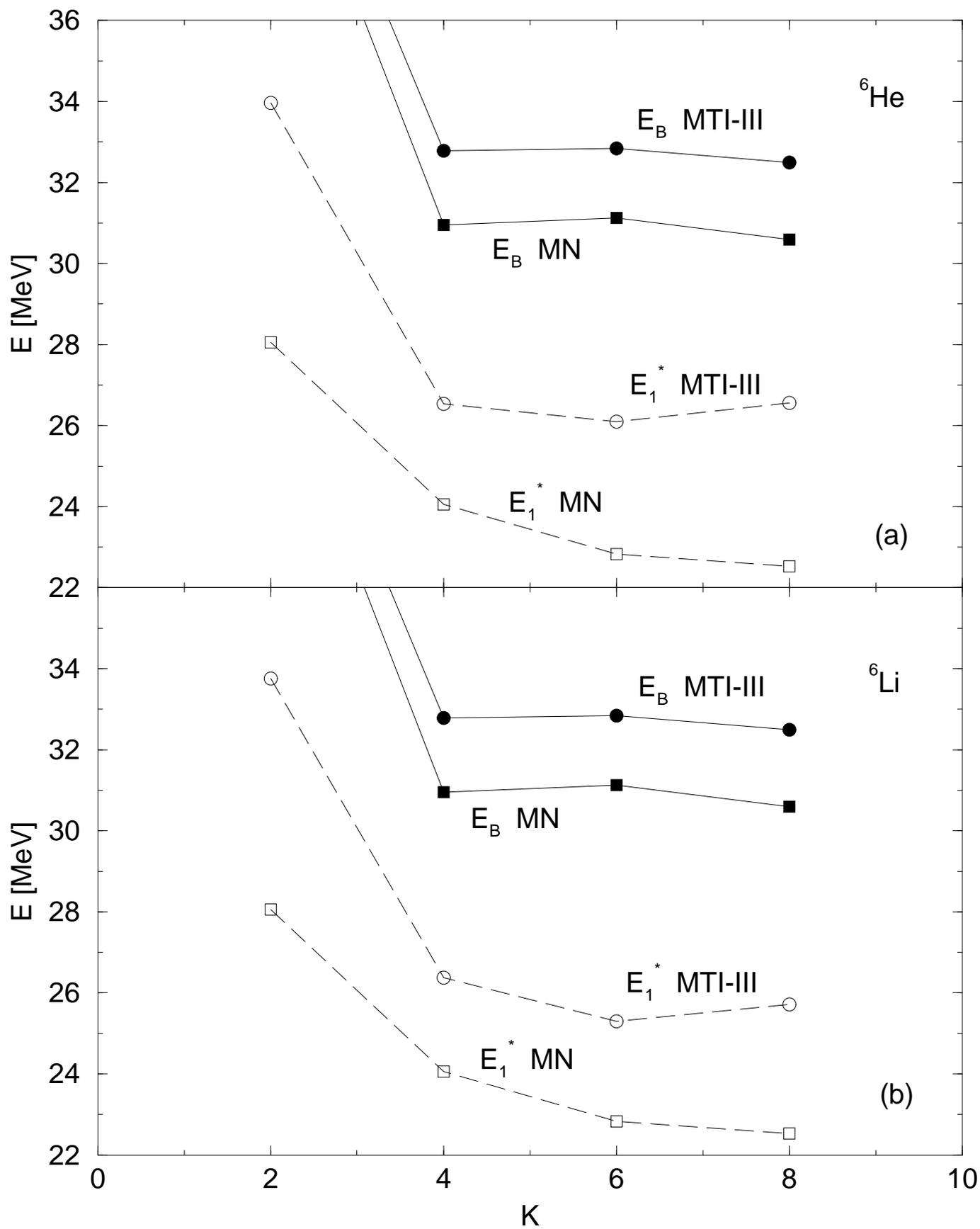
\begin{figure}
\caption{Binding energies $E_B$ and first excitation energies $E_1^\star$ of $^6$He (a) and $^6$Li (b) for  MTI-III [21] 
and MN potentials [22] 
}
\end{figure}

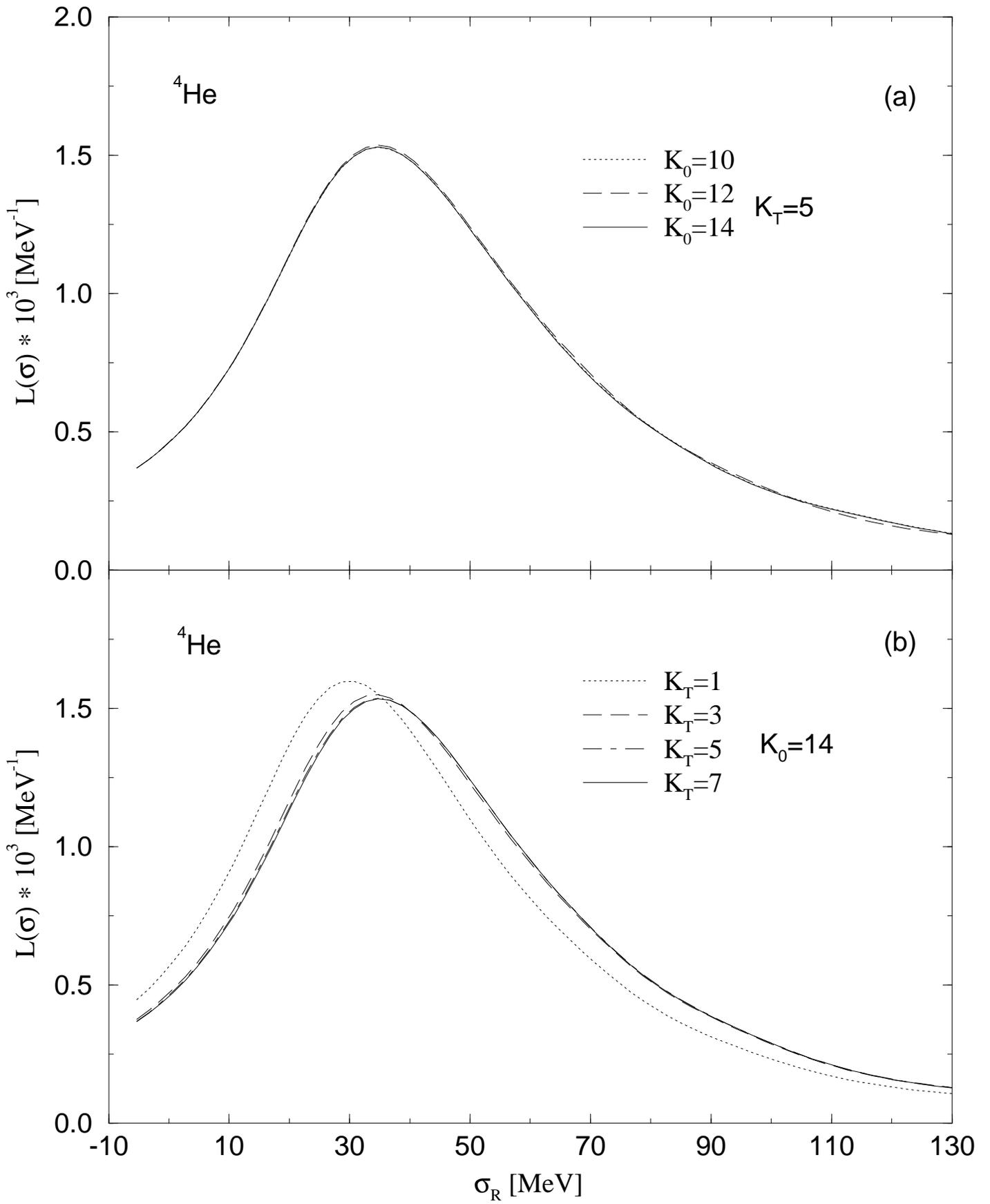
\begin{figure}
\caption{Lorentz integral transform of the $^4$He total photoabsorption 
cross section as a function of $\sigma_R$. In (a) the convergence in the ground
state hyperangular quantum number $K_0$ is shown for a fixed value
of the hyperangular quantum number $K_T$ of the"Lorentz state" $\tilde\Psi$; 
in (b) the convergence in the hyperangular quantum number $K_T$ of the 
"Lorentz state" $\tilde\Psi$ is shown for a fixed value of the ground
state hyperangular quantum number $K_0$.}
\end{figure}

\end{document}